%% file: main.tex
\PassOptionsToPackage{table}{xcolor}
\documentclass[runningheads]{llncs}
\usepackage{amsmath}
\usepackage[ruled,vlined,linesnumbered]{algorithm2e}

\usepackage[T1]{fontenc}
\usepackage{graphicx}
\usepackage{multirow}
\usepackage{booktabs}
\usepackage{pifont}
\usepackage{wasysym}  
\usepackage{tikz}
\usepackage{tabularx}
\usepackage{graphicx}
\usepackage{siunitx}
\usepackage{subfig}   
\usepackage{xcolor}
\usepackage{xspace}
\usepackage{circledtext}
\newcommand{\framework}{\textsc{\textbf{GiAnt}}\xspace}

\newcommand{\mycommentstyle}[1]{\tcp*[r]{\color{gray}\textit{#1}}}

\usepackage{tcolorbox}
\newcommand{\find}[1]{
\begin{tcolorbox}[leftrule=0.5mm,rightrule=0.5mm, toprule=0.5mm,bottomrule=0.5mm,left=2pt,right=2pt,top=2pt,bottom=2pt]
\em #1
\end{tcolorbox}
}

\usepackage[table]{xcolor} 
\usepackage{makecell} 

\newcommand{\cmark}{\textcolor{green!60!black}{\ding{51}}} 
\newcommand{\xmark}{\textcolor{red!80!black}{\ding{55}}}  

\definecolor{distred}{RGB}{237, 28, 36}
\definecolor{distorange}{RGB}{255, 127, 39}
\definecolor{distblue}{RGB}{0, 0, 255}
\definecolor{distbrown}{RGB}{140, 73, 38}
\definecolor{distdarkgreen}{RGB}{117, 189, 66}
\newcommand{\distbar}[5]{%
  \begin{tikzpicture}[x=8pt, y=0.12pt, baseline=-1pt] 
    \fill[distorange] (0,0) rectangle (0.8, #1);
    \fill[distred] (1,0) rectangle (1.8, #2);
    \fill[distblue] (2,0) rectangle (2.8, #3);
    \fill[distdarkgreen] (3,0) rectangle (3.8, #4);
    \fill[distbrown] (4,0) rectangle (4.8, #5);
    \draw[gray!30, ultra thin] (-0.2,0) -- (5,0);
  \end{tikzpicture}%
}

\begin{document}
\title{On the Shoulders of Giants: Empowering Automated Smart Contract Auditing via the GiAnt Corpus}
\titlerunning{Empowering Automated Smart Contract Auditing via the GiAnt Corpus}
\author{Xiaoting Zhang\inst{1,2} \and
Zhipeng Gao\inst{1,2}\thanks{Corresponding author.} \and
Yiran Lv\inst{1,2} \and
Xing Hu\inst{1,2} \and
Feifei Niu\inst{3} \and
Xin Xia\inst{1,2}\protect\footnotemark[1]}
\authorrunning{X. Zhang et al.}
\institute{Zhejiang University, Hangzhou, China \and
Hangzhou High-Tech Zone (Binjiang) Institute of Blockchain and Data Security, Hangzhou, China \and
University of Ottawa, Canada \\
\email{\{xiaotingzhang,zhipeng.gao,xinghu\}@zju.edu.cn} \email{yiranlv7@gmail.com}\\
\email{feifeiniu96@gmail.com}  \email{xin.xia@acm.org} }
\maketitle              
\begin{abstract}
\input{0-abstract}
\keywords{Smart Contract Auditing Dataset \and LLM \and Blockchain.}
\end{abstract}

\input{1-introduction}
\input{2-background-motivation}
\input{3-Framework}
\input{4-evaluation}
\input{5-discussion}
\input{6-related-work}
\input{7-Conclusion}

\section*{Acknowledgment}
This research is supported by the National Science Foundation of China (No. 62572322).  
This research is partially sponsored by the CCF-Huawei Populus Grove Fund. 
We also thank the anonymous reviewers for their insightful comments and suggestions. 

\bibliographystyle{splncs04}
\bibliography{main}
%
%
%
%
%

\end{document}

%% file: 0-abstract.tex
High-quality smart contract auditing datasets are crucial for evaluating security tools and advancing smart contract security research.
Two major limitations of existing datasets are the manual-induced scalability bottleneck and the deficiency in data granularity and diversity.
To address these limitations, we propose \framework (\textbf{\underline{G}}PT-ass\textbf{\underline{i}}sted \textbf{\underline{A}}uditing Dataset Co\textbf{\underline{n}}s\textbf{\underline{t}}ruction), an automated framework designed to curate smart contract auditing datasets by distilling vulnerability insights from real-world auditing reports.
\framework employs a divide-and-conquer strategy coupled with the Chain-of-Thought technique to extract structured vulnerability information from Code4rena reports, followed by an LLM-as-a-judge mechanism to perform rigorous quality assurance.
To evaluate \framework's effectiveness, we run it on 388 real-world audit reports and generate the \textbf{GiAnt Corpus} comprising 7,711 vulnerability findings across five severity levels.
Manual assessment of the dataset demonstrates exceptional reliability in information extraction, achieving a mean quality score of $4.76\pm0.37$ (out of 5) with inter-rater agreement $\kappa$ of 0.88.
We further validate the practicality of our dataset by benchmarking 4 state-of-the-art LLMs on vulnerability detection, code summarization, mitigation recommendation, and automated gas optimization tasks, to establish performance baselines, thereby providing a valuable data foundation for future research in automated smart contract auditing.

%% file: 1-introduction.tex
\section{Introduction}
In recent years, the Ethereum blockchain and smart contract technology has accelerated the explosive growth of Decentralized Finance (DeFi), fundamentally revolutionizing traditional paradigms of digital asset management and financial transactions~\cite{zheng2020overview}. 
As tens of billions of dollars in Total Value Locked (TVL) pour into the DeFi ecosystem, smart contracts bearing immense economic value have inevitably become prime targets for malicious attackers, leading to frequent security incidents and staggering financial losses~\cite{defillama2026,xiang2025automating,gao2020checking,ruan2026improving,gao2020deep}. 

According to recent industry security analyses, the decentralized ecosystem suffered 149 documented major security incidents in 2024 alone, culminating in approximately \$1.42 billion in direct economic losses~\cite{web3hackhub2024}. 
Alarmingly, while the overall frequency of incidents slightly decreased to 135 in 2025, the financial devastation surged to an unprecedented \$3.67 billion~\cite{web3hackhub2025}. 
This drastic escalation in financial impact highlights a significant evolution in attack vectors: malicious attackers have largely shifted away from exploiting superficial syntax errors toward weaponizing complex, high-level security threats. 
Specifically, sophisticated exploit patterns such as \textit{Access Control Vulnerabilities, Business Logic Flaws, Price Oracle Manipulation, and Flash Loan Attacks} have established profound dominance, currently occupying the highest ranks in authoritative industry benchmarks like the latest OWASP Smart Contract Top 10~\cite{owasp2026smartcontract}.
By stealthily exploiting these intricate logical flaws, attackers can bypass existing defenses to execute devastating exploits\cite{sun2024gptscan}. 
Therefore, conducting in-depth research targeting these complex business logic vulnerabilities has become an urgent necessity for securing the modern smart contract ecosystem.

To counter these threats, numerous existing studies have focused on detecting and mitigating specific, traditional types of vulnerability (e.g., \textit{reentrancy} and \textit{integer overflows}~\cite{luu2016making,ding2024vulnerability,chen2025numscout}), leading to the development of various automated analysis tools~\cite{feist2019slither,tsankov2018securify,gao2019smartembed,lin2025actaint,chen2024angels,hu2021automating}. 
\textbf{However, real-world smart contracts are plagued by complicated business logic vulnerabilities that remain largely unable to be discovered by these conventional methods}~\cite{zhang2023demystifying,sendner2024large}. 
Thus, the identification of such intricate flaws remains heavily reliant on \textbf{manual code audits} conducted by premier security firms (e.g., OpenZeppelin and Trail of Bits), which is time-consuming and labor-intensive.

In recent years, the rapid evolution of Large Language Models (LLMs), such as ChatGPT~\cite{achiam2023gpt}, Qwen~\cite{bai2023qwen}, and DeepSeek~\cite{liu2024deepseek}, has redefined the landscape of automated software engineering~\cite{xue2025clean,xue2024selfpico,mai2024human,mai2025towards,dai2025less,dai2024mpcoder,dai2026learner,yu2025enhancing,yan2023closer,wang2024just}. 
By leveraging large-scale pre-training, these models exhibit exceptional proficiency in code comprehension and logical reasoning~\cite{lu2021codexglue}.
Specifically, they can capture intricate semantic patterns and functional dependencies by modeling program contexts. 
Consequently, LLMs offer a promising, cost-effective solution for tackling complex smart contract vulnerabilities.
Despite this potential, the field still lacks a comprehensive, multi-tasking dataset that integrates fine-grained vulnerability insights to drive subsequent research in automated smart contract auditing.
Existing real-world smart contract datasets suffer from two critical limitations:
\circledtextset{resize=real} 
\circledtext*[height=1.8ex]{1} \textbf{Scalability challenge.} 
Manually curating datasets is inherently labor-intensive, making it difficult to scale or evolve alongside the rapid growth of smart contracts. 
For example, DAppScan~\cite{Zheng_2024} required 22 participants and 44 person-months to analyze only 1,199 audit reports. 
This heavy reliance on human effort poses a significant barrier to maintaining a large-scale, up-to-date corpus.
\circledtext*[height=1.8ex]{2} \textbf{Limited data granularity and diversity.} 
Existing benchmarks, such as SmartBugs-CURATED~\cite{Durieux_2020}, predominantly focus on tool-detectable vulnerabilities (e.g., Reentrancy) and often rely on simple, heuristic-based labeling (e.g., ScrawlD~\cite{yashavant2022scrawld}) rather than expert insights. 
Hence, they provide only coarse-grained labels, lacking essential auditing dimensions, such as risk severity, impact analysis, and mitigation strategies.

To address these key limitations, we present \framework (\textbf{\underline{G}}PT-ass\textbf{\underline{i}}sted \textbf{\underline{A}}uditing Dataset Co\textbf{\underline{n}}s\textbf{\underline{t}}ruction), an automated framework for constructing comprehensive and high-quality smart contract auditing datasets.
\framework implements an LLM-driven pipeline to automatically extract vulnerability information from real-world audit reports disclosed by Code4rena, significantly reducing the manual effort required for dataset construction. 
While various platforms provide smart contract audit reports, Code4rena~\cite{code4rena} stands out by explicitly offering fine-grained and diverse data that covers the entire auditing lifecycle, including risk severity, impact analysis, mitigation strategies, and gas optimizations.
Therefore, we crawled all publicly disclosed audit reports from the Code4rena platform as raw data. 
To effectively extract information from extensive audit reports, \framework adopts a divide-and-conquer strategy.
It first leverages the reports' inherent hierarchical structure for segmentation, subsequently conducting fine-grained information distillation on each individual chunk via In-Context Learning (ICL) and Chain-of-Thought (CoT) prompting.
This design harnesses the powerful understanding capabilities of LLMs to extract multi-dimensional auditing artifacts—such as vulnerability descriptions, impact assessments, and mitigation strategies—in a highly scalable manner. 
Furthermore, to mitigate the risks of hallucinations and ensure data integrity, \framework incorporates a quality assurance protocol. 
By employing an LLM-as-a-Judge framework, it systematically validates the extracted data across dimensions of correctness, completeness, and consistency, thereby establishing a robust and trustworthy corpus (i.e., \textbf{GiAnt Corpus}) for downstream security research.
The \textbf{GiAnt Corpus} ultimately comprises 7,711 vulnerability findings across five severity levels (i.e., \textit{High Risk, Medium Risk, Low Risk, Non-Critical, and Gas Optimization}).
The manual evaluation results show that \framework exhibits exceptional reliability in information extraction, achieving a mean quality score of $4.76\pm0.37$ (out of 5).
Moreover, to validate the practicality of our dataset, we benchmark four state-of-the-art LLMs on vulnerability detection, code summarization, mitigation recommendation, and automated gas optimization tasks, and establish performance baselines.

In summary, this paper makes the following contributions: 
\begin{itemize}
    \item[$\bullet$] We propose an LLM-driven framework, named \framework, designed to automatically construct smart contract auditing datasets by extracting vulnerability information from real-world audit reports.
    \item[$\bullet$] We construct the \textbf{GiAnt Corpus} comprising 7,711 detailed vulnerability findings across five severity levels, providing an extensive and finely-categorized corpus for smart contract security research. To the best of our knowledge, this is the most comprehensive dataset to date, providing the broadest support for diverse automated smart contract auditing tasks. 
    \item[$\bullet$] We establish performance baselines for four state-of-the-art LLMs across four tasks (i.e., detection, summarization, remediation, and optimization), demonstrating the dataset's versatility in empowering the full-lifecycle research of automated smart contract auditing.
    \item[$\bullet$] We release \framework and the curated \textbf{GiAnt Corpus}~\cite{replication_package}, in order to facilitate other researchers to replicate our study and verify their own ideas.
\end{itemize}

%% file: 2-background-motivation.tex
\section{Background and Motivation}
\subsection{Smart Contract Audit Reports}
Smart contract auditing is a rigorous security assessment process designed to identify vulnerabilities prior to deployment, typically culminating in detailed reports that document attack vectors, impacts, and recommended mitigations~\cite{hedera_audit_learning}. 

Traditional audits primarily rely on automated static analysis tools or private engagements with specialized security firms. 
However, static tools often suffer from high false positive rates and struggle with complex business logic, while private audits are limited by the perspectives of a few individuals. 
This has catalyzed the emergence of crowdsourced audit platforms, notably Code4rena (C4), which leverage a ``competitive auditing'' model.
C4 is a smart contract auditing platform that has facilitated more than 450 codebase audits and the discovery of thousands of unique vulnerabilities in code from leading crypto projects. 
This platform runs competitions in which auditors compete for rewards from a prize pool by reviewing the target codebase to identify and report vulnerabilities.
In this paradigm, global elite security researchers (referred to as Wardens) compete for lucrative bounties, strongly incentivizing them to bypass easily machine-detectable bugs and uncover highly complex business logic and economic model flaws. 
Following a rigorous multi-party validation process by independent judges, Code4rena finally compiles and publishes highly professional and publicly accessible reports that document discovered vulnerabilities alongside their severity classifications and recommended mitigations.
Unlike traditional reports that often provide limited descriptions, these crowdsourced reports offer unprecedented contextual richness, including detailed impact analysis, executable Proof-of-Concepts (PoCs), and verified sponsor mitigations. 

These unique characteristics make Code4rena audit reports the ideal, high-fidelity data source for constructing datasets tailored to evaluate and fine-tune the deep reasoning capabilities of Large Language Models.

\subsection{LLMs for Automated Dataset Construction}
Large Language Models (LLMs) have recently demonstrated remarkable capabilities across a wide spectrum of software engineering tasks~\cite{xue2025clean,liu2023not,sun2024llm4vuln}, particularly in complex code comprehension, logical reasoning, and unstructured text processing. 
By leveraging extensive pre-training on vast corpora of code and natural language, modern LLMs can deeply understand programming semantics and intricate vulnerability patterns, making them highly promising candidates for automating the extraction and structuring of data from technical documents~\cite{liu2024exploring}. 

Currently, constructing large-scale, high-quality smart contract vulnerability datasets relies heavily on manual curation by human experts, an approach that is prohibitively labor-intensive, time-consuming, and highly susceptible to annotation errors~\cite{chen2025forgellmdrivenframeworklargescale}. 
LLMs can fundamentally overcome these limitations by rapidly parsing massive volumes of audit data, automatically extracting critical vulnerability components, and standardizing information at scale. 
This unprecedented efficiency and scalability strongly motivate us to leverage LLMs to develop an advanced, high-fidelity automated construction paradigm for smart contract audit datasets.
Nevertheless, directly deploying LLMs to extract precise vulnerability information from real-world audit reports introduces substantial technical challenges. 
These reports, typically formatted as lengthy, inconsistently structured Markdown or PDF documents, intertwine natural language descriptions with complex code snippets, execution traces, and mitigation strategies. 
Therefore, the inherent susceptibility of LLMs to ``lost in the middle''  effects and probabilistic hallucinations~\cite{chen2025forgellmdrivenframeworklargescale} in such heterogeneous contexts necessitates a rigorous, verifiable pipeline with alignment mechanisms to ensure high data fidelity.

%% file: 3-Framework.tex
\section{Framework}
To construct high-quality, scalable, and evolvable smart contract auditing datasets, we present \framework, an automated framework that leverages LLMs to distill vulnerability insights from real-world audit reports.

\begin{figure}[t]
    \includegraphics[width=\textwidth]{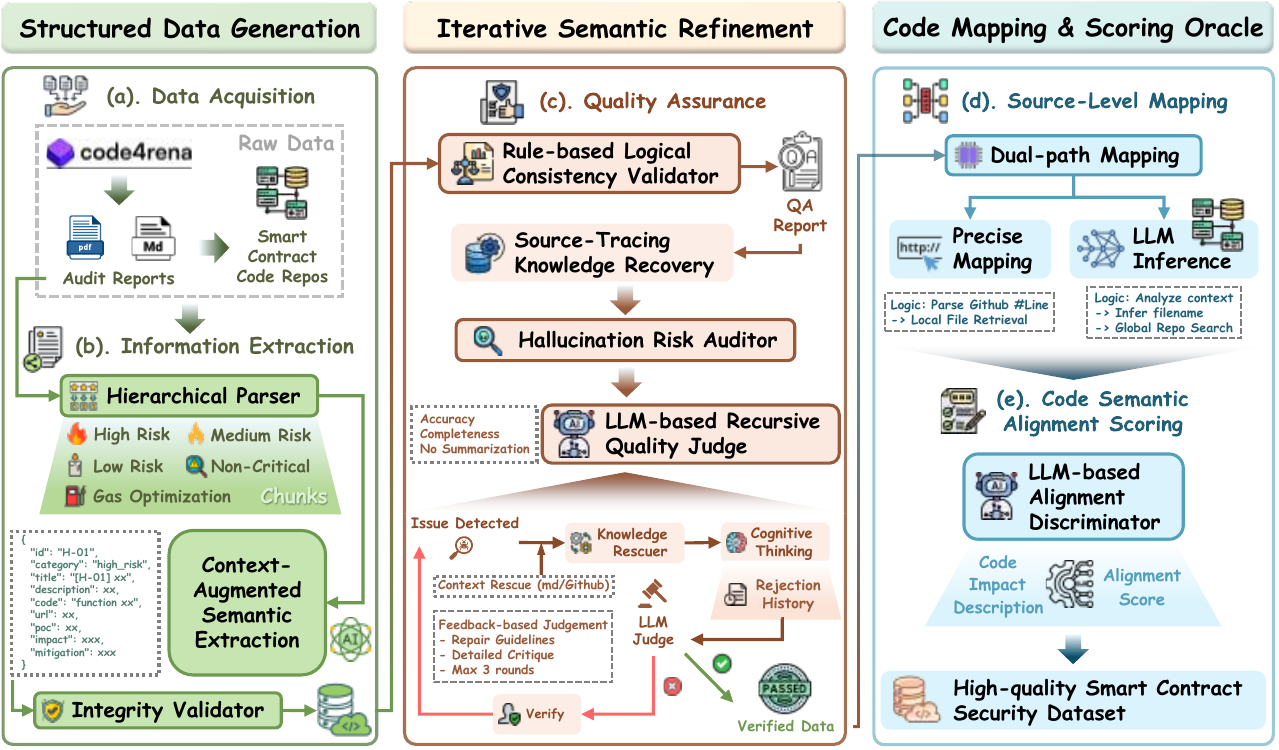}
    \caption{Overview of \framework framework.} 
    \label{fig:overview}
\end{figure}

\subsection{Overview}
The workflow of \framework is outlined in Fig.~\ref{fig:overview},  which consists of three phases. 
First, the \textbf{Structured Data Generation} phase acquires audit reports from Code4rena and extracts essential vulnerability information to construct a preliminary dataset. 
Second, the \textbf{Iterative Semantic Refinement} phase conducts rigorous quality assurance and self-correction on the preliminary dataset to produce a semantically optimized version. 
Finally, the \textbf{Code Mapping \& Scoring Oracle} phase further enhances the code snippets by retrieving source-level contexts and validates these enhancements through strict alignment scoring, ultimately yielding the final High-quality Smart Contract Auditing Dataset.

\subsection{Phase 1: Structured Data Generation}
To transform heterogeneous audit reports and source code into a preliminary dataset, this phase involves data acquisition and information extraction.

\noindent \textbf{(a) Data Acquisition.} 
We started our pipeline by crawling publicly disclosed audit reports from the Code4rena~\cite{code4rena} platform published from 2021 to 2025. 
Objectively, these reports exist in different formats, comprising both Markdown and PDF files. 
Moreover, we clone all corresponding Smart Contract Code Repos that fall within the scope of each audit. 
Since our construction requires both vulnerability information and the corresponding source code, we filter out audit reports where the source code is not publicly accessible. 
The remaining reports and their corresponding repositories constitute our \textit{Raw Data} corpus, which ultimately encompasses 340 distinct projects for further analysis.

\noindent \textbf{(b) Information Extraction.} 
To process this \textit{Raw Data} corpus, we first employ PyMuPDF~\cite{pymupdf,chen2025forgellmdrivenframeworklargescale} to convert all PDF reports into a standardized Markdown format, ensuring structural uniformity for downstream extraction.
Recognizing the inherent hierarchical organization of Code4rena reports—where vulnerabilities are systematically grouped by severity—we implement a regex-based \textbf{hierarchical parser} to segment the documents into semantic \textit{Chunks} (e.g., \textit{High Risk Findings, Medium Risk Findings, Low Risk Findings, Non-Critical Findings, and Gas Optimizations}).
Within each \textit{chunk}, the parser further isolates individual vulnerabilities by matching finding titles, extracting fundamental metadata (such as the \texttt{id} and its associated issue \texttt{url}) and aggregating the remaining text into a \texttt{local\_content} block.
To achieve fine-grained extraction, we subsequently deploy an LLM-based \textbf{Context-Augmented Semantic Extraction} module that leverages each \texttt{local\_content} as its primary input. 
Crucially, we enrich this prompt with external discussions fetched from GitHub issues using the previously extracted \texttt{url}. 
Guided by meticulously crafted prompts, the LLM parses this augmented context to generate a standardized JSON schema, capturing multi-dimensional details including the \textit{Vulnerable Code}, \textit{PoC}, \textit{Impact}, and \textit{Mitigation} strategies.
Finally, to guarantee data robustness, the generated JSON records are immediately routed to an \textbf{Integrity Validator}, which inspects the output for missing fields or invalid, overly brief content. 
Upon detection of anomalies, re-extraction is performed for the deficient components.
The culmination of this phase yields a structured, foundational auditing dataset ready for subsequent semantic refinement.

\subsection{Phase 2: Iterative Semantic Refinement}
This phase focuses on quality assurance to purify the preliminary dataset through the following verification and self-correction mechanisms. 
Firstly, the Logical Consistency Validator identifies missing information to trigger targeted knowledge recovery.
After that, the Hallucination Risk Auditor verifies extraction authenticity. 
Lastly, the Quality Judge conducts semantic evaluation, employing a feedback-driven mechanism to iteratively self-correct flawed records.

\noindent \textbf{(c-1) Logical Consistency Validation.}
Taking the initial dataset built in Phase 1 as input, the \textbf{Logical Consistency Validator} employs a rule-based diagnostic engine to identify extraction omissions by cross-referencing source audit reports with the extracted JSON fields. 
Specifically, it flags missing vulnerability dimensions (e.g., \textit{Mitigation}) when relevant keywords are detected in the source but absent in the dataset. 
These discrepancies are systematically cataloged into a Quality Assurance (QA) Report, which then guides the Knowledge Recovery module to perform a targeted re-extraction of the deficient fragments, ensuring the final corpus faithfully reflects the original expert insights.

\noindent \textbf{(c-2) Hallucination Risk Auditing.}
\framework also employs a \textbf{Hallucination Risk Auditor} to verify the authenticity of the artifacts extracted. 
Since LLMs are strictly prompted to perform extractive parsing rather than abstractive summarization, a faithful output must preserve the exact wording of the original reports. 
Therefore, we utilize a text-inclusion algorithm~\cite{Ji_2023} to compute the lexical overlap rate between the LLM's output and the source fragments.
Any record exhibiting an overlap rate below 0.6 is discarded as a hallucination risk. 
The threshold was experimentally optimized to balance data yield against factual integrity, as preliminary analyses show 0.6 effectively filters non-extractive hallucinations while preserving high-quality samples.
This mechanism ensures that every retained record remains strictly anchored to the ground-truth audit facts, free from LLM-generated noise.

\begin{algorithm}[t]
\caption{Recursive Semantic Repair Loop}
\label{alg:repair}
\small
\KwIn{Deficient record $r$, Source report $R$, Max retries $N=3$}
\KwOut{Verified record $r^*$ or Null}

$C_{loc} \gets \text{Parser}(r.\text{id}, R); \quad C_{ext} \gets \text{Fetch}(r.\text{url})$ \mycommentstyle{Knowledge Rescue}
$C \gets C_{loc} \cup C_{ext}; \quad H \gets [r.\text{init\_reason}]; \quad \text{success} \gets \text{false}$\;

\For{$i \gets 1$ \KwTo $N$}{
    $d_{new} \gets \text{LLM\_Extract}(C, r, H)$ \mycommentstyle{Re-extraction via CoT thinking}
    \If{$d_{new} = \emptyset$}{\textbf{break}}
    
    $(v, \rho) \gets \text{LLM\_Judge}(d_{new}, C)$ \mycommentstyle{$v \in \{\text{PASS, FAIL}\}$; $\rho$: Judgment}
    
    \If{$v = \text{PASS}$}{
        $r \gets d_{new}; \quad \text{success} \gets \text{true}; \quad$ \textbf{break}\;
    }
    $H.\text{append}(\rho); \quad r \gets d_{new}$ \mycommentstyle{Update history with critique $\rho$}
}

\eIf{\text{success}}{\Return $r$}{\text{Discard}($r$); \quad \Return \text{Null} \mycommentstyle{Prune unfixable records}}
\end{algorithm}

\noindent \textbf{(c-3) LLM-based Recursive Quality Judge.}
For deep semantic verification, the extracted records then undergo evaluation by an \textbf{LLM-based Recursive Quality Judge}, which scores the extraction on a 5-point Likert scale~\cite{xia2023empiricalstudysoftwarematerials} against three criteria: \textit{Accuracy} (absence of fabricated content), \textit{Completeness} (preservation of critical code blocks), and \textit{No Summarization} (strict adherence to verbatim extraction).
Crucially, the Judge outputs a definitive PASS/FAIL verdict accompanied by a detailed diagnostic critique.
Records that fail this evaluation are routed into a dynamic \textbf{repair loop}, as detailed in Algorithm~\ref{alg:repair}. 
To equip the Judge with full contextual awareness, a \textit{Knowledge Rescuer} first synthesize local reports $C_{loc}$ and external data $C_{ext}$ into an augmented context $C$ for each deficient record $r$.
Drawing on task-oriented \textit{Chain of Thought}, the system compiles diagnostic critiques $\rho$ into a \textit{Rejection History} $H$ and feeds it back to the extractor. 
Then the LLM is forced into a state of \textit{Cognitive Thinking}, explicitly diagnosing prior shortcomings (e.g., code omissions or over-summarization) before generating a revised output $d_{new}$. 
Each revision is immediately subjected to a real-time self-evaluation by the Judge. 
Successful corrections overwrite the flawed records, while instances that persistently fail after three rounds are permanently discarded. 
Ultimately, this refinement produces highly reliable \textit{Verified Data}, establishing a solid foundation for subsequent phases.

\subsection{Phase 3: Code Mapping \& Scoring Oracle}
The goal of this phase is to turn fragmented vulnerability snippets into complete code samples and verify their quality through two primary steps: Source-Level Code Mapping and Code Semantic Alignment Scoring.

\noindent \textbf{(d) Source-Level Code Mapping.} 
The ``\texttt{vulnerable\_code}'' field extracted from audit reports often lacks structural consistency, manifesting as a mixture of volatile GitHub links and incomplete code fragments (e.g., a single vulnerable code statement).
So, our \framework initiates a \textit{Source-Level Mapping} utilizing a \textbf{Dual-path Mapping} strategy to recover the complete code context. 
For vulnerabilities accompanied by explicit GitHub line anchors (e.g., \texttt{\#L100-L110}), the \textit{Precise Mapping} branch directly parses the URL to retrieve the exact codes. 
Conversely, when explicit links are absent, the pipeline orchestrates an \textit{LLM Inference} mechanism. 
First, the LLM acts as a semantic resolver to extract and verify target filenames (e.g., resolving \texttt{UDA.sol} from a mention of ``fund loss in UDA.sol'') from the vulnerability context (i.e., titles, descriptions, and code fragments). 
\framework then performs a targeted search within identified files; if metadata is absent, it automatically falls back to a global search using the code fragments as anchors to locate the complete logic.
To guarantee the purity of the extracted code, this mapping process enforces strict directory filtering—bypassing noise-inducing folders like \texttt{test}, \texttt{mock}, and \texttt{node\_modules}, thereby ensuring that only the authentic code logic is captured.

\noindent \textbf{(e) Code Semantic Alignment Scoring.} 
In this stage, an \textbf{LLM-based Alignment Discriminator} acts as a senior smart contract auditor to perform cross-verification and evaluate the semantic consistency among the retrieved \textit{Code}, the documented \textit{Impact}, and the \textit{Description}. 
Specifically, the discriminator quantifies the consistency using a 5-point Likert scale: scores of 1--2 represent irrelevant or erroneously extracted snippets; a score of 3 denotes an acceptable extraction that captures the core logic but lacks essential context (e.g., missing modifiers); while scores of 4--5 are exclusively awarded to extractions that demonstrate near-perfect fidelity to the reported description and impact.
By retaining only those elite instances that achieve a score of $\ge$ 4, the final oracle solidifies the generation of our \textit{High-quality Smart Contract Auditing Dataset}.

%% file: 4-evaluation.tex
\section{Evaluation}
In this section, we aim to answer the following four research questions (RQs):
\begin{itemize}
    \item[$\bullet$] \textbf{RQ1: Can \framework effectively construct a high-quality and context-rich smart contract vulnerability dataset from real-world audit reports?} 
    We apply our \framework to 388 real-world audit reports to construct the dataset, subsequently analyzing its statistical characteristics and comparing its comprehensiveness against existing baseline datasets.

    \item[$\bullet$] \textbf{RQ2: How reliable is the vulnerability information extracted by \framework?} 
    To evaluate the consistency of extracted data with expert judgment, we systematically verify the extraction quality by performing a manual evaluation on 400 sampled records across five severity categories.
    
    \item[$\bullet$] \textbf{RQ3: How do existing general LLMs perform across the multi-dimensional auditing tasks facilitated by our dataset?} 
    We benchmark four leading LLMs across pivotal auditing sub-tasks using our dataset to establish a performance baseline, thereby mining the challenges and limitations these models face when confronted with real-world smart contracts.
\end{itemize}

\subsection{RQ1: Effectiveness of \framework}
We begin by running \framework on our previously collected 388 real-world audit reports.
The data construction incurred a total cost of approximately \$597.2 USD, representing a cumulative consumption of 21.3 million tokens at an average rate of 62.6k tokens per report.
Ultimately, our dataset comprises 7,711 multi-level risk findings extracted from the audit reports of 340 distinct projects, with each project's structured data encapsulated in a dedicated JSONL file.

\begin{table}[!htb]
\setlength{\abovecaptionskip}{0pt}
  \setlength{\tabcolsep}{2mm} 
  \centering
  \caption{Data statistics for our \framework corpus and its comparison with existing datasets. JZ: jiuzhou~\cite{zhang2020framework}; SB-C: SmartBugs-CURATED~\cite{Durieux_2020}; SD: ScrawlD~\cite{yashavant2022scrawld}; DAS: DAppScan~\cite{Zheng_2024}; SC-B: SC-Bench~\cite{11028158}; FG: FORGE~\cite{chen2025forgellmdrivenframeworklargescale}.}
  \label{tab:data_statistics}
  \begin{tabular}{l | c c c c c c | c}
    \toprule
    \textbf{Category} & \textbf{JZ} & \textbf{SB-C} & \textbf{SD} & \textbf{DAS} & \textbf{SC-B} & \textbf{FG} & \textbf{GiAnt} \\
    \midrule
    \rowcolor{gray!15} \multicolumn{8}{c}{\textit{Data Curation Method}} \\
    \midrule
    Automated & \LEFTcircle & \LEFTcircle & \CIRCLE & \Circle & \LEFTcircle & \CIRCLE & \CIRCLE \\
    Manual    & \LEFTcircle & \LEFTcircle & \Circle & \CIRCLE & \LEFTcircle & \Circle & \Circle \\
    \midrule
    \rowcolor{gray!15} \multicolumn{8}{c}{\textit{Risk Severity}} \\
    \midrule
    High Risk        & \cmark & \xmark & \xmark & \xmark & \xmark & \cmark & \cmark (1,048) \\
    Medium Risk      & \cmark & \xmark & \xmark & \xmark & \xmark & \cmark & \cmark (2,443) \\
    Low Risk         & \cmark & \xmark & \xmark & \xmark & \xmark & \cmark & \cmark (782) \\
    Non-Critical     & \xmark & \xmark & \xmark & \xmark & \xmark & \cmark & \cmark (723) \\
    Gas Optimization & \xmark & \xmark & \xmark & \xmark & \xmark & \xmark & \cmark (2,715) \\
    \midrule
    \rowcolor{gray!15} \multicolumn{8}{c}{\textit{Key Information Fields}} \\
    \midrule
    Vulnerable Code & \cmark & \cmark & \cmark & \cmark & \cmark & \cmark & \cmark \\
    Description     & \xmark & \xmark & \cmark & \cmark & \cmark & \cmark & \cmark \\
    Severity        & \cmark & \xmark & \xmark & \xmark & \xmark & \cmark & \cmark \\
    Impact          & \xmark & \xmark & \xmark & \xmark & \xmark & \xmark & \cmark \\
    Mitigation      & \cmark & \xmark & \xmark & \xmark & \xmark & \xmark & \cmark \\
    PoC             & \xmark & \xmark & \xmark & \xmark & \xmark & \xmark & \cmark \\
    \midrule
    \rowcolor{gray!15} \multicolumn{8}{c}{\textit{Supported Tasks}} \\
    \midrule
    Vulnerability Detection   & \CIRCLE & \CIRCLE & \CIRCLE & \CIRCLE & \LEFTcircle & \CIRCLE & \CIRCLE \\
    Severity Classification   & \CIRCLE & \Circle & \Circle & \Circle & \Circle & \CIRCLE & \CIRCLE \\
    Code Summarization        & \Circle & \Circle & \Circle & \Circle & \Circle & \CIRCLE & \CIRCLE \\
    Mitigation Generation     & \Circle & \Circle & \Circle & \Circle & \Circle & \Circle & \CIRCLE \\
    Gas Optimization          & \Circle & \Circle & \Circle & \Circle & \Circle & \Circle & \CIRCLE \\
    \bottomrule
     \addlinespace[1ex] 
     \multicolumn{8}{l}{\CIRCLE~Full Support \quad \LEFTcircle~Partial Support \quad \Circle~No Support} \\
  \end{tabular}
\end{table}

Table~\ref{tab:data_statistics} outlines the key statistics of our dataset. 
It comprises a large-scale collection of audited smart contract findings, categorized by risk severity into \textit{High} (1,048), \textit{Medium} (2,443), and \textit{Low} (782) \textit{risks}, alongside 723 \textit{Non-Critical} issues and 2,715 \textit{gas optimization} entries. 
So, compared to existing datasets (e.g., SmartBugs-Wild~\cite{Durieux_2020}, DAppScan~\cite{Zheng_2024} and SC-Bench~\cite{11028158}) that are often curated around specific vulnerability types, \textbf{our real-world sourced dataset exhibits significantly higher categorical diversity and comprehensiveness, providing a more holistic representation of real-world auditing scenarios}.

\begin{figure}[!t]
    \includegraphics[width=\textwidth]{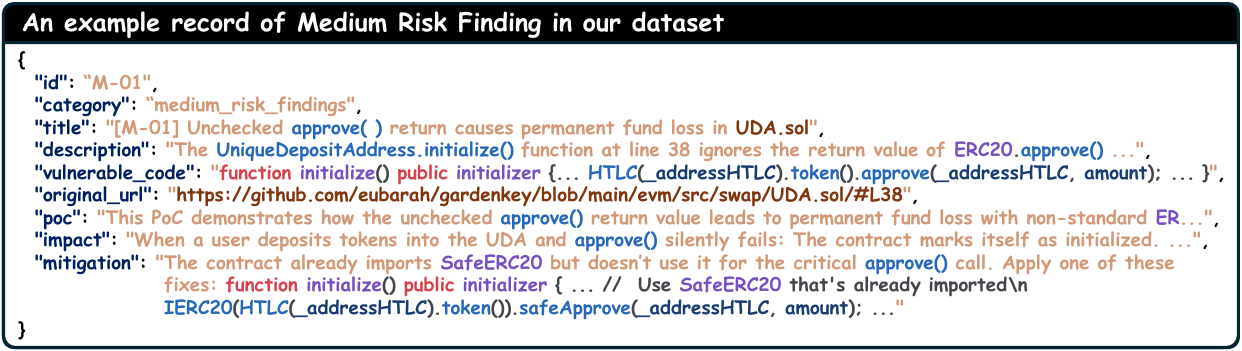}
    \caption{An example record of our dataset in JSON format.} 
    \label{fig:data_example}
\end{figure}

Fig.~\ref{fig:data_example} shows an example of a record in our dataset, illustrating its multi-dimensional structure.
Each entry is comprised of several fields that provide comprehensive information about the audited contract code and its vulnerabilities.
Beyond basic metadata such as a unique identifier (``\texttt{id}'') and classification (``\texttt{category}''), each record provides a ``\texttt{vulnerable\_code}'' field that isolates the specific vulnerability, a ``\texttt{title}'' field that outlines the key vulnerability issues, a ``\texttt{description}'' field that offers expert-level explanation of the underlying logic, and a ``\texttt{original\_url}'' field that documents the code source of the vulnerability. 
Furthermore, each entry is detailed with a ``\texttt{impact}'' field that analyzes potential security consequences, a ``\texttt{poc}'' to demonstrate exploitability, and a ``\texttt{mitigation}'' strategy to provide actionable remediation patches. 
Consequently, \textbf{our dataset offers a substantial leap in information density and functional coverage} over existing benchmarks (e.g., FORGE~\cite{chen2025forgellmdrivenframeworklargescale}), \textbf{enabling a broader spectrum of tasks} ranging from severity classification to automated gas optimization.

\find{\textbf{Answer to RQ1:} Compared to prior works, our dataset built by \framework presents a substantial leap in information density and task-versatility, establishing a new foundation for automated smart contract auditing studies.}

\subsection{RQ2: Reliability of Information Extraction}
\label{sec:quantitative-analysis}
In this section, we aim to assess the reliability of our framework in extracting smart contract vulnerability information. 
To answer this RQ, we conduct a manual analysis to estimate the \framework's extraction quality and assess its consistency with human expert judgments. 

\noindent \textbf{Experiment Setup.}
Given the inherent imbalance in the distribution of finding severities, we employed a \textbf{stratified random sampling} strategy to ensure a representative and statistically significant human evaluation.
Specifically, we randomly sampled 400 records across five severity categories based on their initial distribution from our dataset.
According to Cochran's formula for finite populations~\cite{wohlin2012experimentation}, such a sample size provides a 95\% confidence level with a margin of error of approximately 4.75\%.
The specific sampling breakdown and the resulting scoring distribution for each risk level are detailed in Table~\ref{tab:human_evaluation}.
Subsequently, the first two authors of this paper conducted a manual review of the sampled entries, where 350 records were partitioned equally for independent assessment (175 each) and the remaining 50 were jointly evaluated to measure inter-coder agreement. 
Each entry was scored on a 5-point Likert scale across three dimensions (i.e., categorization accuracy, content integrity, and contextual consistency) by cross-referencing the extracted data with the original audit reports. 

\begin{table} 
\setlength{\abovecaptionskip}{0pt}
  \setlength{\tabcolsep}{2mm} 
  \renewcommand{\arraystretch}{1.5} 
  \centering
  \caption{Stratified Sampling Distribution for Manual Validation}
  \label{tab:human_evaluation}
  \begin{tabular}{lcccc}
    \toprule
      \textbf{Category} & \textbf{Proportion(\%)} & \textbf{Sample Size} & \textbf{\# Score} & \textbf{\# Distri.} \\
    \midrule
    \textit{High Risk} & 13.6 & 54 & $4.83\pm0.22$ & \distbar{0}{0}{4}{9}{87} \\
    \textit{Medium Risk} & 31.7 & 127 & $4.81\pm0.27$ & \distbar{0}{1}{4}{9}{86} \\
    \textit{Low Risk} & 10.1 & 40 & $4.69\pm0.26$ & \distbar{0}{0}{2.5}{25}{72.5} \\
    \textit{Non-Critical} & 9.4 & 38 & $4.78\pm0.24$ & \distbar{0}{0}{2}{16}{82} \\
     \textit{Gas Optimization} & 35.2 & 141 & $4.68\pm0.60$ & \distbar{1}{3}{4}{10}{82} \\
    \midrule
    Total & 100 & 400 & $4.76\pm0.37$ & \distbar{0.25}{1.5}{3.75}{11.75}{82.75} \\
    \bottomrule          
    \end{tabular}                                 
\end{table}

\noindent \textbf{Experiment Result.}
We assessed the objectivity of the manual evaluation using Cohen’s Kappa coefficient on the overlapping 50-item subset, yielding a score of \textbf{$\kappa = 0.88$}, indicating ``near-perfect'' agreement. 
The final evaluation results are presented in Table~\ref{tab:human_evaluation}, which shows that the dataset achieves a \textbf{high average quality score of $4.76\pm0.37$ (out of 5) across all categories}, demonstrating the exceptional precision and consistency of our automated extraction framework in preserving expert insights.
The narrow standard deviation further underscores the remarkable stability and robustness of our \framework in processing diverse audit categories.
Moreover, we also provide a granular distribution of these scores across the five risk categories to illustrate the model's performance consistency across varying levels of vulnerability severity.
The distribution histograms showcase a \textbf{significant concentration of scores at the highest level} (score 5), indicating that the vast majority of parsed samples maintain near-perfect alignment with the original audit reports.

\find{\textbf{Answer to RQ2:} \framework exhibits exceptional reliability, achieving a mean quality score of $4.76\pm0.37$ with near-perfect inter-rater agreement ($\kappa = 0.88$). This high score, coupled with a narrow standard deviation, underscores our \framework’s robustness in preserving expert insights with high fidelity.}

\subsection{RQ3: Benchmarking General LLMs}
Beyond raw code and basic descriptions, our dataset encapsulates a rich spectrum of auditing artifacts, including severity classifications, impact analyses, mitigation strategies, Proof-of-Concept (PoC) exploits and gas optimization insights. 
Therefore, this multi-dimensional information provides a foundational corpus for investigating the entire continuum of smart contract auditing, spanning both security robustness and execution efficiency. 

To demonstrate its utility, in this RQ, we curate specialized subsets from our dataset to evaluate the performance of popular Large Language Models (LLMs) on four pivotal sub-tasks: vulnerability detection, code summarization, mitigation recommendation and automated gas optimization. 
By benchmarking these representative tasks, we aim to explore the potential of our dataset in fueling automated, multi-tasking auditing frameworks.

\textbf{Dataset Setup.}
To construct the evaluation benchmark, we randomly selected 386 samples covering four risk tiers from our vulnerability dataset.
Such a sample size is statistically significant, ensuring a margin of error of $\pm5$\% at a 95\% confidence level.
Given the specific focus on gas efficiency, we separately sampled an additional 386 instances for the gas optimization task.

\textbf{Model Selection.}
We conducted our experiments across four LLMs (i.e., GPT-3.5-turbo, GPT-4o, Qwen3-Coder-480B-A35B-Instruct and DeepSeek-V3).
All models were configured with a zero-decoding temperature to facilitate deterministic benchmarking.
For automated assessment, we implemented an LLM-as-a-Judge framework utilizing GPT-4o, whereby a structured scoring rubric was employed to objectively quantify the accuracy of the generated content.

\textbf{Metrics.}
For vulnerability detection, we formulated the task as a classification problem and employed metrics including Accuracy (Acc.), Precision (Pre.), Recall (Rec.), and macro-averaged F1-score (F1.). 
For generative tasks (i.e., Code Summarization, Mitigation Recommendation, and Gas Optimization), we utilized BERTScore~\cite{zhang2019bertscore} (\#BERTS) to quantify the semantic alignment between LLM-generated insights and ground truth reports. 
Besides, we conducted an expert-level assessment (\#LLMS) using an LLM-judge, which rated the samples on a 5-point Likert scale based on their relevance and consistency.
These results were reported as ``mean $\pm$ standard deviation'' and supplemented by distribution bar charts (\#Distri.) to provide a visualization of data consistency.

\subsubsection{Vulnerability Detection Task.}
As the foundational step in automated smart contract auditing, the vulnerability detection task is formulated as a binary or multi-class classification problem. 
Specifically, it requires the LLM to identify the presence of exploitable vulnerabilities in the given contract and subsequently categorize them into specific severity tiers (e.g., High, Medium, Low or Non-Critical Risk).
In this section, we aim to benchmark the detection capabilities of several LLMs on our dataset, thereby revealing their proficiency and limitations in diagnosing complex, real-world vulnerabilities.

\begin{table} 
\setlength{\abovecaptionskip}{0pt}
  \setlength{\tabcolsep}{1.5mm} 
  \renewcommand{\arraystretch}{1.2} 
  \centering
  \caption{Vulnerability Detection Performance of Different LLMs}
  \label{tab:bug_detection}
  \begin{tabular}{lcccccccc}
    \toprule
\multirow{2}{*}{\textbf{Model}} & \multicolumn{4}{c}{\textbf{Binary Classification}} & \multicolumn{4}{c}{\textbf{Multi-class Classification}} \\
    \cmidrule(lr){2-5} \cmidrule(lr){6-9}
      & \textbf{Acc.} & \textbf{Pre.} & \textbf{Rec.} & \textbf{F1.} & \textbf{Acc.} & \textbf{Pre.} & \textbf{Rec.} & \textbf{F1.} \\
    \midrule
    \textit{GPT-3.5-turbo} & 0.613 & 0.918 & 0.625 & 0.744 & 0.375 & 0.330 & 0.334 & 0.301 \\
    \textit{GPT-4o} & 0.813 & 0.918 & 0.869 & 0.893 & 0.363 & 0.421 & 0.320 & 0.284 \\
    \textit{Qwen3-Coder-Instruct} & 0.673 & 0.907 & 0.708 & 0.796 & 0.370 & 0.318 & 0.296 & 0.280 \\
    \textit{DeepSeek-V3} & 0.745 & 0.908 & 0.797 & 0.849 & 0.365 & 0.371 & 0.328 & 0.302 \\
    \bottomrule          
    \end{tabular}                                 
\end{table}

\begin{figure}[!htb] 
    \centering
    \subfloat[Performance of GPT-3.5-turbo]{%
       \includegraphics[width=0.45\textwidth]{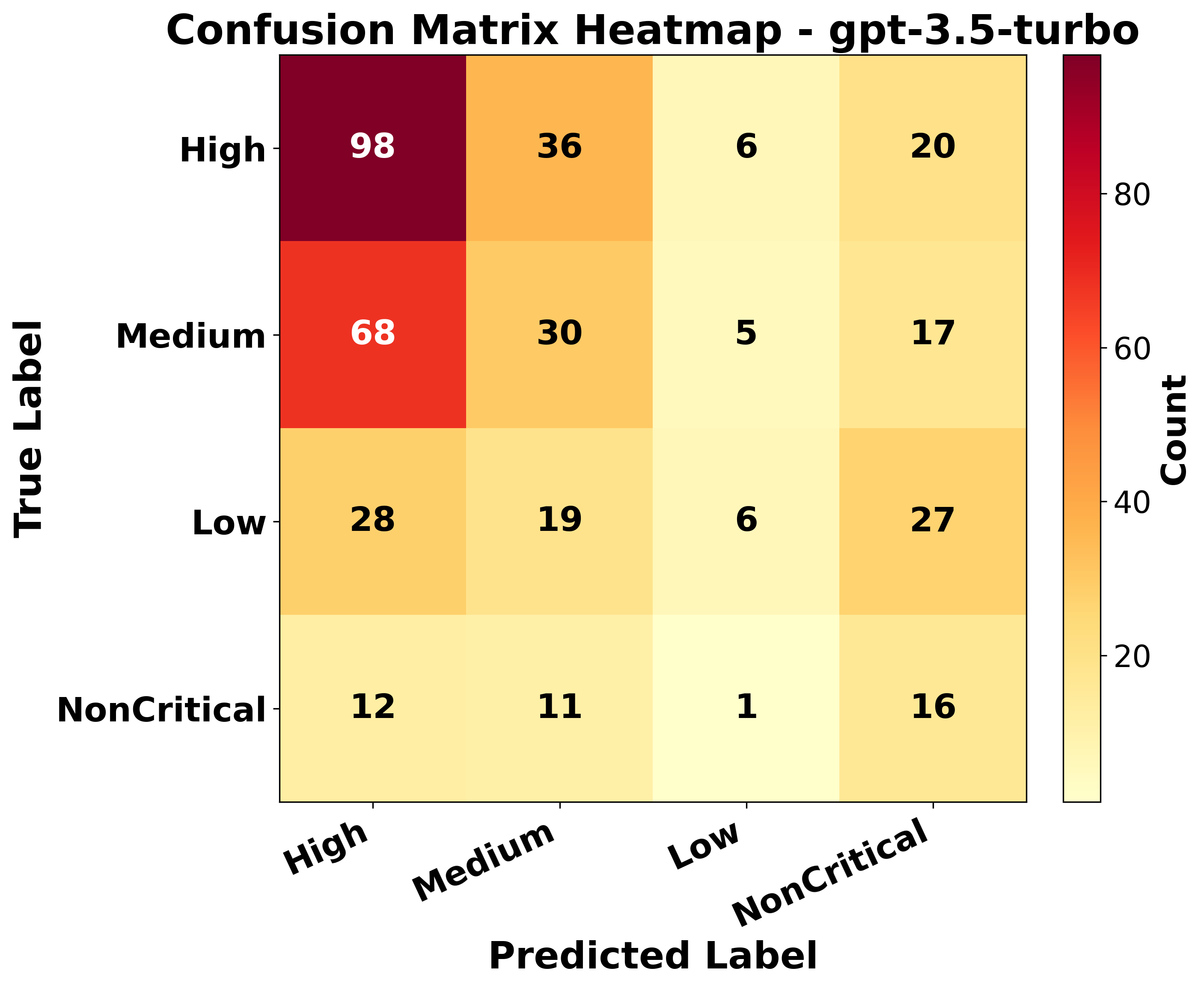}\label{fig:gpt3.5}} 
    \qquad
    \subfloat[Performance of GPT-4o]{%
       \includegraphics[width=0.45\textwidth]{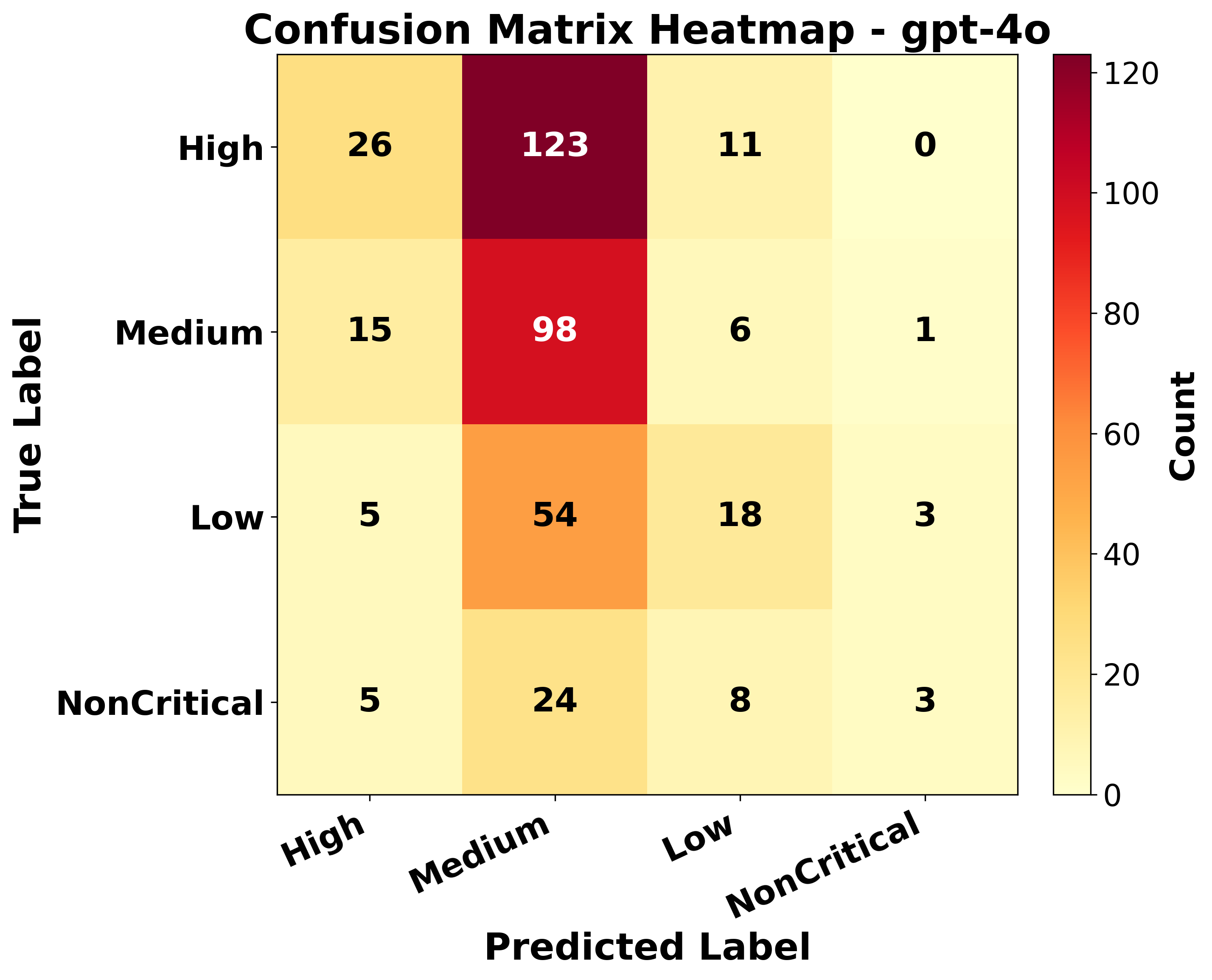}\label{fig:gpt4o}}  
    \vspace{1em} 
    \\ 
    \subfloat[Performance of Qwen3-Coder]{%
       \includegraphics[width=0.45\textwidth]{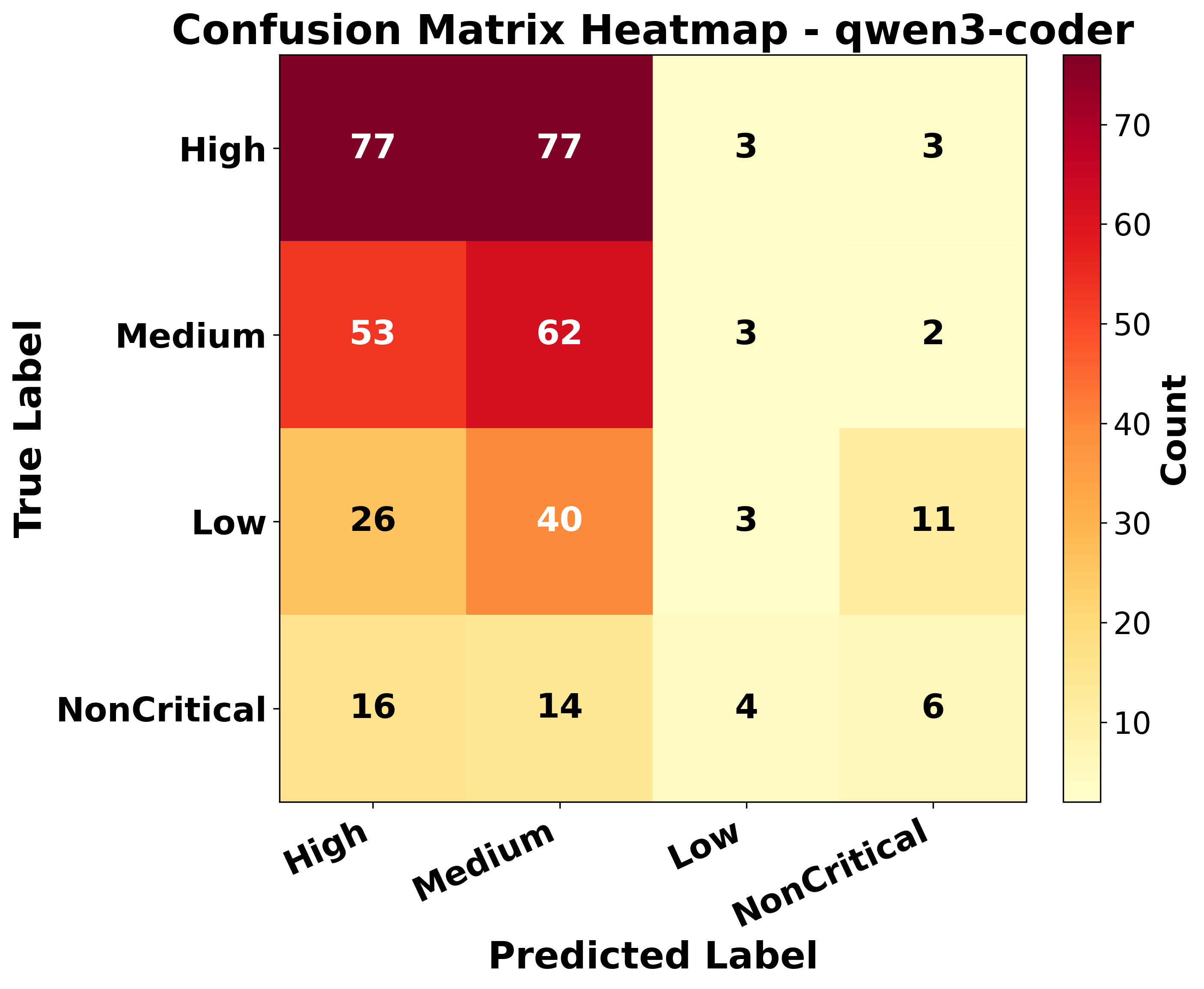}\label{fig:qwen}} 
    \qquad
    \subfloat[Performance of DeepSeek-V3]{%
       \includegraphics[width=0.45\textwidth]{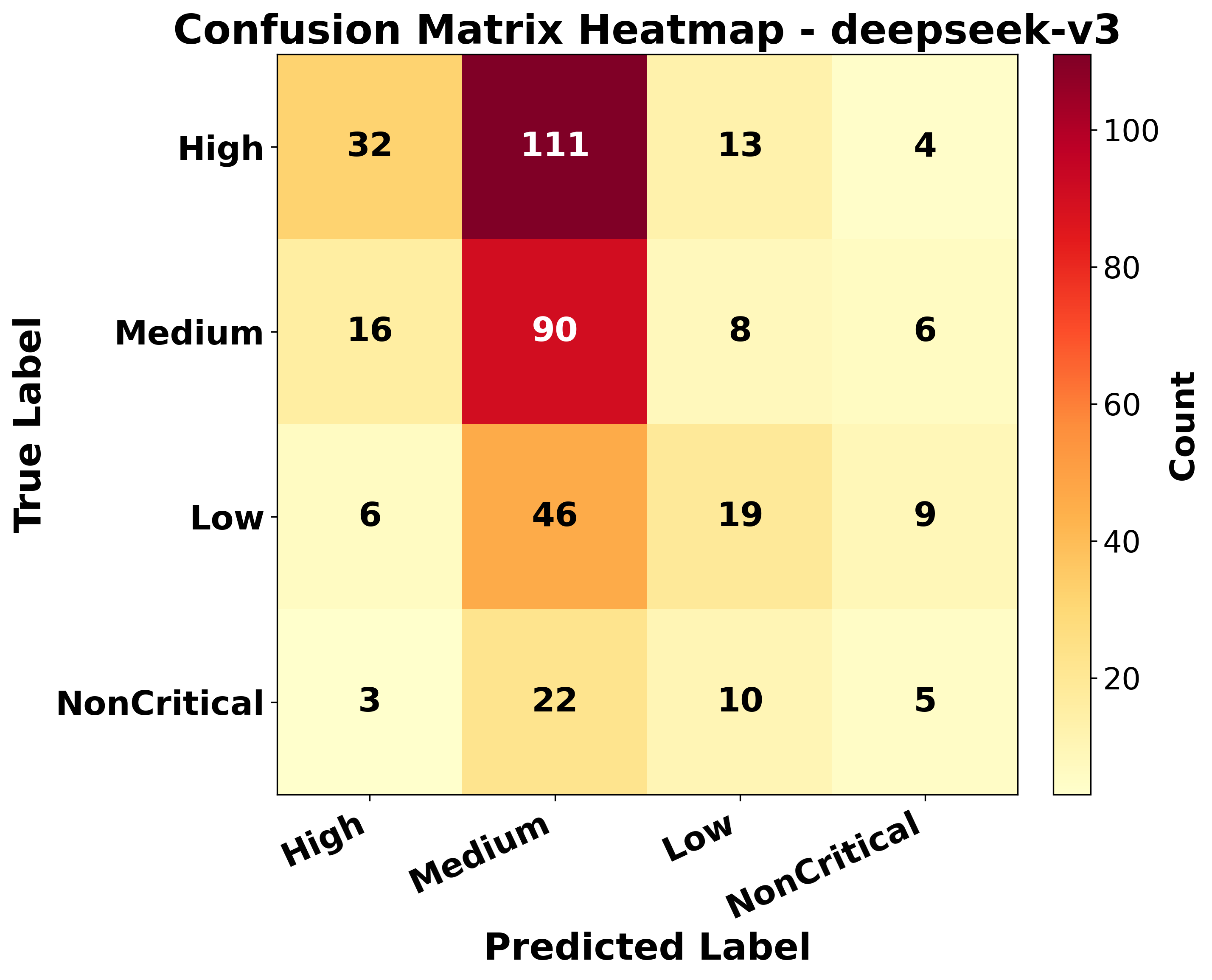}\label{fig:deepseek}} 
    \caption{Confusion matrices of LLMs for multi-class vulnerability prediction.}
    \label{fig:confusion_matrix}
\end{figure}

\noindent \textbf{Experiment Result.}
Table~\ref{tab:bug_detection} lists the vulnerability detection performance of four LLMs in binary and multi-class classification settings.
The experimental results show that while all models demonstrate robust security awareness in binary classification — with GPT-4o leading at a 0.893 F1-score — performance significantly falters in multi-class severity categorization, where F1-scores drop to a range of 0.280–0.302. 
This disparity indicates that while LLMs are effective at identifying the presence of bugs, they struggle to assess their potential impact. 
Furthermore, to analyze the models' misclassification trends across different severity tiers, we visualize the multi-class classification results through confusion matrix heatmaps (Fig.~\ref{fig:confusion_matrix}).
The confusion matrices further demonstrate that the LLMs exhibit a high misclassification rate in assessing actual vulnerability risks, with a notable tendency toward ``over-warning''. 
This is evidenced by the frequent misclassification of lower-risk vulnerabilities into higher severity categories.
For instance, as shown in the heatmaps of GPT-4o and DeepSeek-V3, a significant portion of Low and Non-Critical instances are erroneously flagged as Medium or High risk.
Therefore, simple prompt engineering is insufficient for precise vulnerability severity assessment, highlighting the need for more advanced techniques such as fine-tuning to achieve expert-level risk judgment~\cite{10.1145/3795692}.

\subsubsection{Code Summarization Task.}
In the professional auditing workflow, a deep comprehension of a contract's business logic is a prerequisite for identifying security flaws.
Similarly, the effectiveness of an automated auditor hinges on its ability to decode the semantic intent behind raw code. 
To evaluate this foundational capability, we benchmark four LLMs on the code summarization task, where models are required to generate natural language explanations of given code snippets through reasoning. 
Beyond general descriptions, we further task the models with generating an impact analysis of potential logical risks, thereby evaluating their capacity for semantic reasoning and threat assessment.

\begin{table}
\setlength{\abovecaptionskip}{0pt}
  \renewcommand{\arraystretch}{1.5} 
  \centering
  \caption{Code Summarization Performance of Different LLMs}
  \label{tab:code_summarization}
  \resizebox{\textwidth}{!}{
  \begin{tabular}{lcccccc}
    \toprule
\multirow{2}{*}{\textbf{Model}} & \multicolumn{3}{c}{\textbf{Code Understanding}} & \multicolumn{3}{c}{\textbf{Risk Analysis}} \\
    \cmidrule(lr){2-4} \cmidrule(lr){5-7}
      & \#BERTS & \#LLMS & \#Distri. & \#BERTS & \#LLMS & \#Distri. \\
    \midrule
    \textit{GPT-3.5-turbo} & 0.740 & $1.97\pm0.90$ & \distbar{34.25}{40.75}{20.5}{3}{1.5} & 0.734 & $2.64\pm0.96$ & \distbar{8.75}{40}{34.75}{11.75}{4.75} \\
    \textit{GPT-4o} & 0.758 & $2.51\pm1.09$ & \distbar{17.25}{37.5}{28.75}{10}{6.5} & 0.737 & $2.97\pm1.06$ & \distbar{5.25}{31.5}{35}{18}{10.25} \\
    \textit{Qwen3-Coder-Instruct} & 0.757 & $2.48\pm1.13$ & \distbar{18.75}{39}{25.5}{9}{7.75} & 0.733 & $2.86\pm0.98$ & \distbar{4.5}{35.5}{37}{15.75}{7.25} \\
    \textit{DeepSeek-V3} & 0.747 & $2.40\pm1.11$ & \distbar{22}{38.25}{23.5}{10.5}{5.75} & 0.735 & $2.89\pm1.05$ & \distbar{7}{31.5}{35}{18.25}{8.25} \\
    \bottomrule          
    \end{tabular}  
    }
\end{table}

\noindent \textbf{Experiment Result.}
As shown in Table~\ref{tab:code_summarization}, the evaluated LLMs exhibit a notable discrepancy between automated semantic metrics and expert evaluations.
While all models achieve relatively high BERTScore values (exceeding 0.733), indicating strong semantic alignment with ground-truth reports, the LLM-judge ratings consistently fall below the 3.0 threshold across both code understanding and impact analysis tasks. 
This disparity suggests that despite generating semantically plausible summaries, current models struggle to achieve the granular, domain-specific comprehension required for professional smart contract auditing. 
Consequently, these results highlight an urgent need to integrate more specialized augmentation techniques to bridge the gap between basic semantic mimicking and genuine deep logic understanding in the smart contract domain.

\subsubsection{Mitigation Recommendation Task.}
In the smart contract security lifecycle, vulnerability mitigation is one of the most critical activities. 
Mitigation recommendation refers to the generation of secure code patches to remediate identified vulnerabilities without introducing new security risks. 
Recent studies have attempted to use LLMs to perform the automated program repair (APR) task for smart contracts. 
The core idea is to provide the vulnerable code and its contextual description to the model, and then leverage the model's semantic reasoning and code generation capabilities to output a corrected implementation.

\begin{table} 
\setlength{\abovecaptionskip}{0pt}
  \setlength{\tabcolsep}{3mm} 
  \renewcommand{\arraystretch}{1.5} 
  \centering
  \caption{Mitigation Recommendation Performance of Different LLMs}
  \label{tab:mitigation}
  \begin{tabular}{lccc}
    \toprule
      \textbf{Model} & \textbf{\# BERTS} & \textbf{\# LLMS} & \textbf{\# Distri.} \\
    \midrule
    \textit{GPT-3.5-turbo} & 0.769 & $3.64\pm1.05$ & \distbar{0.5}{10.75}{44.25}{12.75}{31.75} \\
    \textit{GPT-4o} & 0.778 & $4.03\pm1.00$ & \distbar{0.25}{4}{34.75}{14.75}{46.25} \\
    \textit{Qwen3-Coder-Instruct} & 0.780 & $4.01\pm0.98$ & \distbar{0}{4.75}{33}{18.25}{44} \\
    \textit{DeepSeek-V3} & 0.782 & $4.09\pm1.02$ & \distbar{0}{6.5}{28.75}{14.25}{50.5} \\
    \bottomrule          
    \end{tabular}                                 
\end{table}

\noindent \textbf{Experiment Result.}
Table~\ref{tab:mitigation} presents the performance of the evaluated LLMs in generating remediation strategies for identified vulnerabilities.
We can see that all evaluated LLMs demonstrate a strong capacity for mitigation recommendation, achieving high semantic alignment with human experts (\#BERTS $\geq$ 0.769, \#LLMS $\geq$ 3.64). 
These results suggest that when provided with detailed vulnerability descriptions, LLMs can effectively leverage their inherent reasoning to formulate technically plausible mitigation patches. 
However, the significant standard deviation—approximating 1.0—across all ratings reveals a lack of output stability, underscoring the necessity of integrating advanced techniques to enhance the reliability and precision of automated auditing frameworks.

\subsubsection{Automated Gas Optimization Task.}
Gas optimization is a critical pillar of smart contract auditing, as inefficient code execution directly translates to prohibitive transaction costs and potential Denial-of-Service (DoS) vulnerabilities in resource-constrained blockchain environments. 
Unlike traditional software optimization, it requires a balance between logical correctness and the minimization of EVM (Ethereum Virtual Machine) opcodes. 
Recently, the advent of LLMs has catalyzed a shift towards automated gas optimization, where models are tasked with performing end-to-end code refactoring. 
By taking raw Solidity code as input and leveraging their profound understanding of code semantics, LLMs aim to generate functionally equivalent yet gas-efficient variants.

\begin{table} 
\setlength{\abovecaptionskip}{0pt}
  \setlength{\tabcolsep}{3mm} 
  \renewcommand{\arraystretch}{1.5} 
  \centering
  \caption{Gas Optimization Performance of Different LLMs}
  \label{tab:gas_optimization}
  \begin{tabular}{lccc}
    \toprule
    \textbf{Model} & \textbf{\# BERTS} & \textbf{\# LLMS} & \textbf{\# Distri.} \\
    \midrule
    \textit{GPT-3.5-turbo} & 0.715 & $2.80\pm1.18$ & \distbar{6}{46.67}{26}{4.33}{17}  \\
    \textit{GPT-4o} & 0.739 & $3.19\pm1.21$ & \distbar{4}{28.67}{36.67}{5.67}{25} \\
    \textit{Qwen3-Coder-Instruct} & 0.737 & $3.27\pm1.18$ & \distbar{3.33}{24.67}{38.67}{8.67}{24.67} \\
    \textit{DeepSeek-V3} & 0.736 & $3.38\pm1.24$ & \distbar{3}{25}{34}{7.33}{30.67} \\
    \bottomrule          
    \end{tabular}                                 
\end{table}

\noindent \textbf{Experiment Result.}
Table~\ref{tab:gas_optimization} summarizes the gas optimization performance of four evaluated LLMs.
The results demonstrate that all evaluated models can generate functionally equivalent, gas-efficient variants with high semantic alignment to expert solutions, as evidenced by BERTScore values exceeding 0.715 and LLM-judge ratings nearing the median (i.e., approximating 3.0). 
These findings suggest that LLMs possess a foundational capacity for code refactoring and resource optimization through profound semantic understanding. 
However, the substantial standard deviation of approximately 1.2 indicates significant instability in output quality, highlighting a performance volatility that underscores the critical need for integrating more sophisticated techniques to enhance LLM precision and reliability in automated gas optimization.

\find{\textbf{Answer to RQ3:} While LLMs establish a foundational baseline across all tasks, zero-shot prompting is insufficient for professional auditing, necessitating the integration of advanced techniques to enhance precision and reliability.}

%% file: 5-discussion.tex
\section{Threats to Validity}

\noindent \textbf{Internal Validity.} 
The reliance on LLM-based extraction and evaluation may introduce inherent inaccuracies and biases, potentially overlooking domain-specific hallucinations in complex smart contract logic. 
However, this risk is significantly mitigated by two factors: 
First, our manual evaluation in Section~\ref{sec:quantitative-analysis} shows that the extracted artifacts maintain high precision and consistency with the original report. 
Second, our dataset will be publicly hosted on GitHub, enabling community-driven verification and continuous refinement via the issue tracking system, which ensures the long-term data reliability for practical applications.

\noindent \textbf{External Validity.} 
The first threat concerns selection bias, as our corpus is exclusively curated from Code4rena. 
However, this approach actually strengthens data authenticity since the reports consist of vulnerabilities discovered in real-world projects and verified by multiple independent security experts, thereby bolstering ecological validity rather than diminishing it.
The second threat involves generalizability constraints, as our pipeline leverages Code4rena's specific layout, potentially limiting its direct applicability to other platforms. 
Nevertheless, the core underlying artifacts (e.g., vulnerability descriptions and impacts) are universal. 
Extending our framework to unstructured reports only requires specific pre-processing modules without altering our fundamental methodology.
The third threat relates to model dependency from our selection of GPT-4o.
This is mitigated by our extraction-centric objective and modular architecture: GPT-4o is tasked with verbatim information extraction within short contexts—a capability well within its current stable limits—rather than complex reasoning. 
This design ensures that our methodology remains robust across different models, allowing for straightforward substitution as more advanced LLMs emerge.

%% file: 6-related-work.tex
\section{Related Work}
Constructing high-quality audit datasets is pivotal for smart contract security. 

\noindent \textbf{Manual Curation.}
Early benchmarks primarily rely on rigorous manual curation to ensure data quality. 
For instance, SmartBugs-CURATED~\cite{Durieux_2020} provides high-quality, manual-tagged labels for 69 contracts, while DAppSCAN~\cite{Zheng_2024} involves extensive human labor (44 person-months) to process audit reports. 
While these datasets offer expert-level insights, their prohibitive manual overhead severely limits their scalability and prevents them from capturing the rapidly evolving vulnerability landscape in the smart contract ecosystem.

\noindent \textbf{Automated Construction.}
To improve scalability, researchers have proposed automated labeling using static analysis tools. 
ScrawlD~\cite{yashavant2022scrawld}, for example, employs a majority-voting mechanism among multiple tools. 
However, such datasets suffer from a ``performance ceiling'': the quality of the dataset is inherently capped by the underlying tools' limitations. 
Furthermore, they provide only binary or coarse-grained labels (e.g., jiuzhou~\cite{zhang2020framework} and SC-Bench~\cite{11028158}), lacking the multi-dimensional audit insights (e.g., detailed analyses and mitigation strategies) required for complex reasoning tasks.
Recently, the advent of LLMs has inspired automated dataset construction. 
For instance, FORGE~\cite{chen2025forgellmdrivenframeworklargescale} employs an LLM-driven pipeline to automatically extract and classify over 27k vulnerabilities into the CWE hierarchy, validating the potential of LLMs for information extraction. 

\noindent \textbf{Our Solution.}
In this study, we present \framework to automatically extract high-quality vulnerability artifacts from real-world audit reports, greatly reducing the manual effort required for dataset curation. 
By leveraging the elite crowdsourced intelligence of Code4rena, we further constructed an unprecedented, comprehensive multi-task dataset (i.e., the \textbf{GiAnt Corpus}) tailored for automated smart contract auditing research, which comprises 7,711 vulnerability findings across five severity levels.
Unlike prior datasets confined to basic vulnerability detection, our corpus enables comprehensive evaluation of LLMs across a spectrum of advanced tasks, ranging from vulnerability detection to mitigation generation.

%% file: 7-Conclusion.tex
\section{Conclusion}
In this paper, we propose \framework, an automated framework for curating high-quality smart contract audit datasets from real-world reports. 
By leveraging a divide-and-conquer strategy coupled with Chain-of-Thought reasoning, \framework systematically extracts complex vulnerabilities while employing an LLM-as-a-Judge mechanism to mitigate hallucinations and ensure data quality. 
As a result, \framework has produced a comprehensive dataset (i.e., the \textbf{GiAnt Corpus}) comprising 7,711 verified findings across five severities. 
We further utilized this corpus to establish a performance baseline for state-of-the-art LLMs across critical auditing dimensions, including detection, summarization, remediation, and gas optimization. 
By bridging the gap between audit reports and actionable research artifacts, \textbf{GiAnt} provides a foundational corpus that enables multi-task automated smart contract auditing studies for both academia and industry.

%% file: main.bib
@inproceedings{Durieux_2020, series={ICSE ’20},
   title={Empirical review of automated analysis tools on 47,587 Ethereum smart contracts},
   DOI={10.1145/3377811.3380364},
   booktitle={Proceedings of the ACM/IEEE 42nd International Conference on Software Engineering},
   author={Durieux, Thomas and Ferreira, João F. and Abreu, Rui and Cruz, Pedro},
   collection={ICSE ’20} 
}

@article{Zheng_2024,
   title={DAppSCAN: Building Large-Scale Datasets for Smart Contract Weaknesses in DApp Projects},
   volume={50},
   ISSN={2326-3881},
   DOI={10.1109/tse.2024.3383422},
   number={6},
   journal={IEEE Transactions on Software Engineering},
   author={Zheng, Zibin and Su, Jianzhong and Chen, Jiachi and Lo, David and Zhong, Zhijie and Ye, Mingxi},
   year={2024}
}

@misc{chen2025forgellmdrivenframeworklargescale,
      title={FORGE: An LLM-driven Framework for Large-Scale Smart Contract Vulnerability Dataset Construction}, 
      author={Jiachi Chen and Yiming Shen and Jiashuo Zhang and Zihao Li and John Grundy and Zhenzhe Shao and Yanlin Wang and Jiashui Wang and Ting Chen and Zibin Zheng},
      year={2025},
      eprint={2506.18795},
      archivePrefix={arXiv},
      primaryClass={cs.CR},
      url={https://arxiv.org/abs/2506.18795}
}

@INPROCEEDINGS{11028158,
  author={Xia, Shihao and He, Mengting and Song, Linhai and Zhang, Yiying},
  booktitle={2025 IEEE/ACM International Workshop on Large Language Models for Code (LLM4Code)}, 
  title={SC-Bench: A Large-Scale Dataset for Smart Contract Auditing}, 
  year={2025},
  volume={},
  number={},
  pages={57-64},
  doi={10.1109/LLM4Code66737.2025.00012}
}

@techreport{web3hackhub2024,
  author = {{SolidityScan}},
  title = {{Web3HackHub} 2024 Annual Security Report: Analyzing 149 Incidents and \$1.42B Losses},
  institution = {SolidityScan},
  year = {2024},
  url = {https://solidityscan.com/}
}

@misc{owasp2026smartcontract,
  author = {{OWASP Foundation}},
  title = {{OWASP} Smart Contract Top 10 - 2026},
  howpublished = {\url{https://owasp.org/www-project-smart-contract-top-10/}}
}

@misc{web3hackhub2025,
  author = {{SolidityScan}},
  title = {{Web3HackHub}: 2025 {Web3} Security Incidents Statistics},
  howpublished = {\url{https://solidityscan.com/web3hackhub?year=2025}},
  note = {Accessed: 2026-03-02}
}

@article{zheng2020overview,
  title={An overview on smart contracts: Challenges, advances and platforms},
  author={Zheng, Zibin and Xie, Shaoan and Dai, Hong Ning and Chen, Weili and Chen, Xiangping and Weng, Jian and Imran, Muhammad},
  journal={Future Generation Computer Systems},
  volume={105},
  pages={475--491},
  year={2020},
  publisher={Elsevier}
}

@misc{defillama2026,
  author = {{DeFiLlama}},
  year = {2026},
  howpublished = {\url{https://defillama.com/}}
}

@inproceedings{luu2016making,
  title={Making smart contracts smarter},
  author={Luu, Loi and Chu, Duc-Hiep and Olickel, Hrishi and Saxena, Prateek and Hobor, Aquinas},
  booktitle={Proceedings of the 2016 ACM SIGSAC conference on computer and communications security},
  pages={254--269},
  year={2016}
}

@inproceedings{tsankov2018securify,
  title={Securify: Practical security analysis of smart contracts},
  author={Tsankov, Petar and Dan, Andrei and Drachsler-Cohen, Dana and Gervais, Arthur and Buenzli, Florian and Vechev, Martin},
  booktitle={Proceedings of the 2018 ACM SIGSAC conference on computer and communications security},
  year={2018}
}

@inproceedings{feist2019slither,
  title={Slither: a static analysis framework for smart contracts},
  author={Feist, Josselin and Grieco, Gustavo and Groce, Alex},
  booktitle={2019 IEEE/ACM 2nd international workshop on emerging trends in software engineering for blockchain (WETSEB)},
  pages={8--15},
  year={2019},
  organization={IEEE}
}

@article{chen2025numscout,
  title={Numscout: Unveiling numerical defects in smart contracts using llm-pruning symbolic execution},
  author={Chen, Jiachi and Shao, Zhenzhe and Yang, Shuo and Shen, Yiming and Wang, Yanlin and Chen, Ting and Shan, Zhenyu and Zheng, Zibin},
  journal={IEEE Transactions on Software Engineering},
  year={2025},
  publisher={IEEE}
}

@article{ding2024vulnerability,
  title={Vulnerability detection with code language models: How far are we?},
  author={Ding, Yangruibo and Fu, Yanjun and Ibrahim, Omniyyah and Sitawarin, Chawin and Chen, Xinyun and Alomair, Basel and Wagner, David and Ray, Baishakhi and Chen, Yizheng},
  journal={arXiv preprint arXiv:2403.18624},
  year={2024}
}

@inproceedings{zhang2023demystifying,
  title={Demystifying exploitable bugs in smart contracts},
  author={Zhang, Zhuo and Zhang, Brian and Xu, Wen and Lin, Zhiqiang},
  booktitle={2023 IEEE/ACM 45th International Conference on Software Engineering (ICSE)},
  pages={615--627},
  year={2023},
  organization={IEEE}
}

@inproceedings{sendner2024large,
  title={Large-scale study of vulnerability scanners for Ethereum smart contracts},
  author={Sendner, Christoph and Petzi, Lukas and Stang, Jasper and Dmitrienko, Alexandra},
  booktitle={2024 IEEE Symposium on Security and Privacy (SP)},
  pages={2273--2290},
  year={2024},
  organization={IEEE}
}

@inproceedings{sun2024gptscan,
  title={Gptscan: Detecting logic vulnerabilities in smart contracts by combining gpt with program analysis},
  author={Sun, Yuqiang and Wu, Daoyuan and Xue, Yue and Liu, Han and Wang, Haijun and Xu, Zhengzi and Xie, Xiaofei and Liu, Yang},
  booktitle={Proceedings of the IEEE/ACM 46th international conference on software engineering},
  pages={1--13},
  year={2024}
}

@misc{code4rena,
  author = {{Code4rena}},
  howpublished = {\url{https://code4rena.com/}}
}

@misc{hedera_audit_learning,
  author = {{Hedera Hashgraph, LLC}},
  title = {What Is a Smart Contract Audit?},
  howpublished = {\url{https://hedera.com/learning/smart-contract-audit/}}
}

@inproceedings{liu2023not,
  title={Not the end of story: An evaluation of ChatGPT-driven vulnerability description mappings},
  author={Liu, Xin and Tan, Yuan and Xiao, Zhenghang and Zhuge, Jianwei and Zhou, Rui},
  booktitle={Findings of the Association for Computational Linguistics: ACL 2023},
  pages={3724--3731},
  year={2023}
}

@article{sun2024llm4vuln,
  title={Llm4vuln: A unified evaluation framework for decoupling and enhancing llms' vulnerability reasoning},
  author={Sun, Yuqiang and Wu, Daoyuan and Xue, Yue and Liu, Han and Ma, Wei and Zhang, Lyuye and Liu, Yang and Li, Yingjiu},
  journal={arXiv:2401.16185},
  year={2024}
}

@inproceedings{liu2024exploring,
  title={Exploring $\{$ChatGPT's$\}$ capabilities on vulnerability management},
  author={Liu, Peiyu and Liu, Junming and Fu, Lirong and Lu, Kangjie and Xia, Yifan and Zhang, Xuhong and Chen, Wenzhi and Weng, Haiqin and Ji, Shouling and Wang, Wenhai},
  booktitle={33rd USENIX Security Symposium (USENIX Security 24)},
  pages={811--828},
  year={2024}
}

@misc{pymupdf,
  title = {PyMuPDF},
  howpublished = {\url{https://github.com/pymupdf/PyMuPDF}},
}

@article{zhang2019bertscore,
  title={Bertscore: Evaluating text generation with bert},
  author={Zhang, Tianyi and Kishore, Varsha and Wu, Felix and Weinberger, Kilian Q and Artzi, Yoav},
  journal={arXiv preprint arXiv:1904.09675},
  year={2019}
}

@book{wohlin2012experimentation,
  title={Experimentation in software engineering},
  author={Wohlin, Claes and Runeson, Per and H{\"o}st, Martin and Ohlsson, Magnus C and Regnell, Bj{\"o}rn and Wessl{\'e}n, Anders and others},
  volume={236},
  year={2012},
  publisher={Springer}
}

@article{achiam2023gpt,
  title={Gpt-4 technical report},
  author={Achiam, Josh and Adler, Steven and Agarwal, Sandhini and Ahmad, Lama and Akkaya, Ilge and Aleman, Florencia Leoni and Almeida, Diogo and Altenschmidt, Janko and Altman, Sam and Anadkat, Shyamal and others},
  journal={arXiv:2303.08774},
  year={2023}
}

@article{liu2024deepseek,
  title={Deepseek-v3 technical report},
  author={Liu, Aixin and Feng, Bei and Xue, Bing and Wang, Bingxuan and Wu, Bochao and Lu, Chengda and Zhao, Chenggang and Deng, Chengqi and Zhang, Chenyu and Ruan, Chong and others},
  journal={arXiv:2412.19437},
  year={2024}
}

@article{bai2023qwen,
  title={Qwen technical report},
  author={Bai, Jinze and Bai, Shuai and Chu, Yunfei and Cui, Zeyu and Dang, Kai and Deng, Xiaodong and Fan, Yang and Ge, Wenbin and Han, Yu and Huang, Fei and others},
  journal={arXiv preprint arXiv:2309.16609},
  year={2023}
}

@article{lu2021codexglue,
  title={Codexglue: A machine learning benchmark dataset for code understanding and generation},
  author={Lu, Shuai and Guo, Daya and Ren, Shuo and Huang, Junjie and Svyatkovskiy, Alexey and Blanco, Ambrosio and Clement, Colin and Drain, Dawn and Jiang, Daxin and Tang, Duyu and others},
  journal={arXiv:2102.04664},
  year={2021}
}

@inproceedings{zhang2020framework,
  title={A framework and dataset for bugs in ethereum smart contracts},
  author={Zhang, Pengcheng and Xiao, Feng and Luo, Xiapu},
  booktitle={2020 IEEE international conference on software maintenance and evolution (ICSME)},
  pages={139--150},
  year={2020},
  organization={IEEE}
}

@article{yashavant2022scrawld,
  title={Scrawld: A dataset of real world ethereum smart contracts labelled with vulnerabilities},
  author={Yashavant, Chavhan Sujeet and Kumar, Saurabh and Karkare, Amey},
  journal={arXiv:2202.11409},
  year={2022}
}

@misc{xia2023empiricalstudysoftwarematerials,
      title={An Empirical Study on Software Bill of Materials: Where We Stand and the Road Ahead}, 
      author={Boming Xia and Tingting Bi and Zhenchang Xing and Qinghua Lu and Liming Zhu},
      year={2023},
      eprint={2301.05362},
      archivePrefix={arXiv},
      primaryClass={cs.SE},
      url={https://arxiv.org/abs/2301.05362}, 
}

@article{ruan2026improving,
  title={Improving Gas Efficiency in Smart Contracts: Data-Driven Insights and LLM-Assisted Remediation},
  author={Ruan, Yijie and Gao, Zhipeng and Chen, Jiachi and Bao, Lingfeng and Yang, Xiaohu},
  journal={IEEE Transactions on Software Engineering},
  year={2026},
  publisher={IEEE}
}

@inproceedings{yan2023closer,
  title={A closer look at different difficulty levels code generation abilities of chatgpt},
  author={Yan, Dapeng and Gao, Zhipeng and Liu, Zhiming},
  booktitle={2023 38th IEEE/ACM International Conference on Automated Software Engineering (ASE)},
  pages={1887--1898},
  year={2023},
  organization={IEEE}
}

@article{dai2026learner,
  title={Learner-Tailored Program Repair: A Solution Generator with Iterative Edit-Driven Retrieval Enhancement},
  author={Dai, Zhenlong and Zhao, Zhuoluo and Wang, Hengning and Tang, Xiu and Wu, Sai and Yao, Chang and Gao, Zhipeng and Chen, Jingyuan},
  journal={arXiv preprint arXiv:2601.08545},
  year={2026}
}

@article{wang2024just,
  title={Just-in-time todo-missed commits detection},
  author={Wang, Haoye and Gao, Zhipeng and Hu, Xing and Lo, David and Grundy, John and Wang, Xinyu},
  journal={IEEE Transactions on Software Engineering},
  volume={50},
  number={11},
  pages={2732--2752},
  year={2024},
  publisher={IEEE}
}

@inproceedings{hu2021automating,
  title={Automating user notice generation for smart contract functions},
  author={Hu, Xing and Gao, Zhipeng and Xia, Xin and Lo, David and Yang, Xiaohu},
  booktitle={2021 36th IEEE/ACM International Conference on Automated Software Engineering (ASE)},
  pages={5--17},
  year={2021},
  organization={IEEE}
}

@inproceedings{gao2020deep,
  title={When deep learning meets smart contracts},
  author={Gao, Zhipeng},
  booktitle={Proceedings of the 35th IEEE/ACM international conference on automated software engineering},
  pages={1400--1402},
  year={2020}
}

@inproceedings{mai2025towards,
  title={Towards Better Answers: Automated Stack Overflow Post Updating},
  author={Mai, Yubo and Gao, Zhipeng and Wang, Haoye and Bi, Tingting and Hu, Xing and Xia, Xin and Sun, Jianling},
  booktitle={2025 IEEE/ACM 47th International Conference on Software Engineering (ICSE)},
  pages={591--603},
  year={2025},
  organization={IEEE}
}

@inproceedings{dai2024mpcoder,
  title={Mpcoder: Multi-user personalized code generator with explicit and implicit style representation learning},
  author={Dai, Zhenlong and Yao, Chang and Han, WenKang and Yuanying, Yuanying and Gao, Zhipeng and Chen, Jingyuan},
  booktitle={Proceedings of the 62nd Annual Meeting of the Association for Computational Linguistics (Volume 1: Long Papers)},
  pages={3765--3780},
  year={2024}
}

@article{yu2025enhancing,
  title={Enhancing Domain-Specific Code Completion via Collaborative Inference with Large and Small Language Models},
  author={Yu, Jingrong and Gao, Zhipeng and Bao, Lingfeng and Liu, Zhongxin},
  journal={ACM Transactions on Software Engineering and Methodology},
  year={2025},
  publisher={ACM New York, NY}
}

@inproceedings{dai2025less,
  title={Less is more: Adaptive program repair with bug localization and preference learning},
  author={Dai, Zhenlong and Chen, Bingrui and Zhao, Zhuoluo and Tang, Xiu and Wu, Sai and Yao, Chang and Gao, Zhipeng and Chen, Jingyuan},
  booktitle={Proceedings of the AAAI Conference on Artificial Intelligence},
  volume={39},
  number={1},
  pages={128--136},
  year={2025}
}

@article{mai2024human,
  title={Are human rules necessary? generating reusable apis with cot reasoning and in-context learning},
  author={Mai, Yubo and Gao, Zhipeng and Hu, Xing and Bao, Lingfeng and Liu, Yu and Sun, JianLing},
  journal={Proceedings of the ACM on Software Engineering},
  volume={1},
  number={FSE},
  pages={2355--2377},
  year={2024},
  publisher={ACM New York, NY, USA}
}

@inproceedings{xue2024selfpico,
  title={Selfpico: Self-guided partial code execution with llms},
  author={Xue, Zhipeng and Gao, Zhipeng and Wang, Shaohua and Hu, Xing and Xia, Xin and Li, Shanping},
  booktitle={Proceedings of the 33rd ACM SIGSOFT International Symposium on Software Testing and Analysis},
  pages={1389--1401},
  year={2024}
}

@article{xue2025clean,
  title={Clean Code, Better Models: Enhancing LLM Performance with Smell-Cleaned Dataset},
  author={Xue, Zhipeng and Zhang, Xiaoting and Gao, Zhipeng and Hu, Xing and Gao, Shan and Xia, Xin and Li, Shanping},
  journal={ACM Transactions on Software Engineering and Methodology},
  year={2025},
  publisher={ACM New York, NY}
}

@article{chen2024angels,
  title={Angels or demons: investigating and detecting decentralized financial traps on ethereum smart contracts},
  author={Chen, Jiachi and Hu, Jiang and Xia, Xin and Lo, David and Grundy, John and Gao, Zhipeng and Chen, Ting},
  journal={Automated Software Engineering},
  volume={31},
  number={2},
  pages={63},
  year={2024},
  publisher={Springer}
}

@inproceedings{lin2025actaint,
  title={ACTaint: Agent-Based Taint Analysis for Access Control Vulnerabilities in Smart Contracts},
  author={Lin, Huarui and Gao, Zhipeng and Chen, Jiachi and Chen, Xiang and Yang, Xiaohu and Bao, Lingfeng},
  booktitle={2025 40th IEEE/ACM International Conference on Automated Software Engineering (ASE)},
  pages={2555--2567},
  year={2025},
  organization={IEEE}
}

@article{xiang2025automating,
  title={Automating comment generation for smart contract from bytecode},
  author={Xiang, Jianhang and Gao, Zhipeng and Bao, Lingfeng and Hu, Xing and Chen, Jiayuan and Xia, Xin},
  journal={ACM Transactions on Software Engineering and Methodology},
  volume={34},
  number={3},
  pages={1--31},
  year={2025},
  publisher={ACM}
}

@article{Ji_2023,
   title={Survey of Hallucination in Natural Language Generation},
   volume={55},
   ISSN={1557-7341},
   DOI={10.1145/3571730},
   number={12},
   journal={ACM Computing Surveys},
   publisher={Association for Computing Machinery (ACM)},
   author={Ji, Ziwei and Lee, Nayeon and Frieske, Rita and Yu, Tiezheng and Su, Dan and Xu, Yan and Ishii, Etsuko and Bang, Ye Jin and Madotto, Andrea and Fung, Pascale},
   year={2023}
}

@article{gao2020checking,
  title={Checking smart contracts with structural code embedding},
  author={Gao, Zhipeng and Jiang, Lingxiao and Xia, Xin and Lo, David and Grundy, John},
  journal={IEEE Transactions on Software Engineering},
  volume={47},
  number={12},
  pages={2874--2891},
  year={2020},
  publisher={IEEE}
}

@inproceedings{gao2019smartembed,
  title={Smartembed: A tool for clone and bug detection in smart contracts through structural code embedding},
  author={Gao, Zhipeng and Jayasundara, Vinoj and Jiang, Lingxiao and Xia, Xin and Lo, David and Grundy, John},
  booktitle={2019 IEEE International Conference on Software Maintenance and Evolution (ICSME)},
  pages={394--397},
  year={2019},
  organization={IEEE}
}

@misc{replication_package,
  title = {Our replication package},
  year = {2026},
  howpublished = {https://zenodo.org/records/19325553}
}

@article{10.1145/3795692,
author = {Li, Xiaoqi and Li, Zongwei and Li, Wenkai and Zhang, Yuqing and Wang, Xin},
title = {No More Hidden Pitfalls? Exposing Smart Contract Bad Practices with LLM-Powered Hybrid Analysis},
year = {2026},
publisher = {Association for Computing Machinery},
address = {New York, NY, USA},
issn = {1049-331X},
doi = {10.1145/3795692},
note = {Just Accepted},
journal = {ACM Trans. Softw. Eng. Methodol.}
}
